\documentclass{elsart41}
\usepackage{graphics}
\usepackage{graphicx}
\usepackage{amssymb}

\begin{document}

\begin{frontmatter}



\title{Many-polaron problem by cluster perturbation theory}


\author[Gr,Er]{M. Hohenadler\corauthref{Name1}},
\ead{hohenadler@itp.tugraz.at}
\author[Er]{G. Wellein},
\author[Gw]{A. Alvermann},\,
\author[Gw]{H. Fehske}

\address[Gr]{Institute for Theoretical and Computational Physics, TU Graz, 8010 Graz, Austria}
\address[Er]{Computing Center, University Erlangen, 91058 Erlangen, Germany}
\address[Gw]{Institute for Physics, Ernst-Moritz-Arndt University Greifswald, 17487 Greifswald, Germany}

\corauth[Name1]{Corresponding author. Tel: +43 316 873-8189, Fax: ext. 8677}

\begin{abstract}
The carrier-density dependence of the photoemission spectrum of the Holstein
many-polaron model is studied using cluster perturbation theory
combined with an improved cluster diagonalization by Chebychev expansion. 
\end{abstract}

\end{frontmatter}


Polaronic quasiparticles originating from strong electron-phonon interaction
play an important role in polar solids, including alkali halides, transition
metal oxides/perovskites and, in particular, quasi one-dimensional (1d)
materials such as MX chains, conjugated polymers or organic charge-transfer
complexes. However, the theoretical description of the underlying models
represents a challenging open problem.

Recently, significant effects of finite carrier density on the spectral
properties of a polaronic system have been discovered for the spinless
Holstein model \cite{HoNevdLWeLoFe04}, revealing the shortcomings of widely
used single-polaron theories. The Hamiltonian of the 1d tight-binding ($t$)
Holstein model reads
\[
  H
  =
  -t \sum_{\langle i,j\rangle} c^\dag_i c^{\phantom{\dag}}_j
  +\omega_0\sum_i b^\dag_i b^{\phantom{\dag}}_i
  -\sqrt{E_\mathrm{p}\omega_0} \sum_i \hat{n}_i (b^\dag_i + b^{\phantom{\dag}}_i)
  .
\]
Here $c^\dag_i$ ($b^\dag_i$) creates a spinless fermion (a phonon) at lattice
site $i$, and $\hat{n}_i=c^\dag_i c^{\phantom{\dag}}_i$. The parameters are
the adiabaticity ratio $\omega_0/t$, and the dimensionless coupling constant
$\lambda=E_\mathrm{p}/2t$, with the polaron binding energy $E_\mathrm{p}$.

While existing work based on finite-cluster calculations was restricted
either in energy or momentum resolution \cite{HoNevdLWeLoFe04,CaGrSt99},
the cluster perturbation theory (CPT, see \cite{HoAivdL03} and references
therein) yields accurate results in the thermodynamic limit.

Here, we calculate the zero-temperature Green function $G(k,\omega) =
\langle\langle c^{\phantom{\dag}}_k;c_k^\dag\rangle\rangle_\omega$ using CPT
in combination with a Chebychev expansion technique
\cite{WeWeAlFe05}. The latter does not suffer from the shortcomings of other
methods, and exploits the analytical separation of the symmetric phonon mode
\cite{SyHuBeWeFe04} in order to ensure well-converged results. Calculations
have been done on parallel supercomputers, with the largest matrix dimension
exceeding $10^{10}$. Within CPT, the one-particle Green function is obtained
from $G^{-1} = G^{(\mathrm{c})-1}-V$, where $G^{(\mathrm{c})}$
and $V$ denote the corresponding real-space cluster Green function and the
intercluster hopping matrix, respectively, and subsequent Fourier
transformation \cite{HoAivdL03}.

\begin{figure*}[th!]
    \centering
    \includegraphics[width=0.47\textwidth]{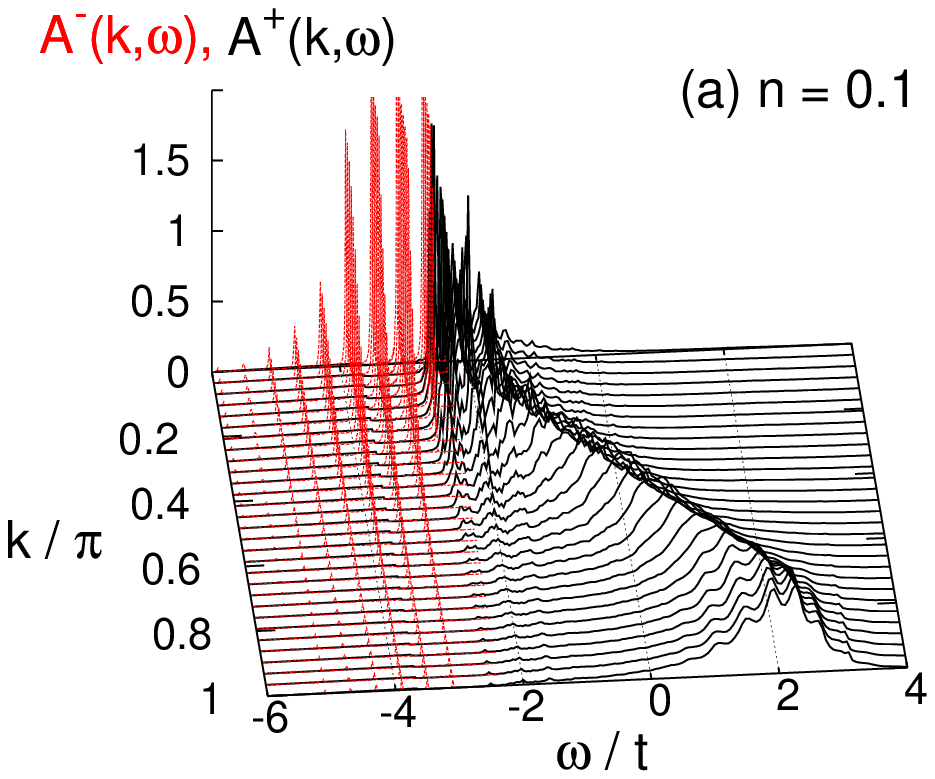}
    \includegraphics[width=0.47\textwidth]{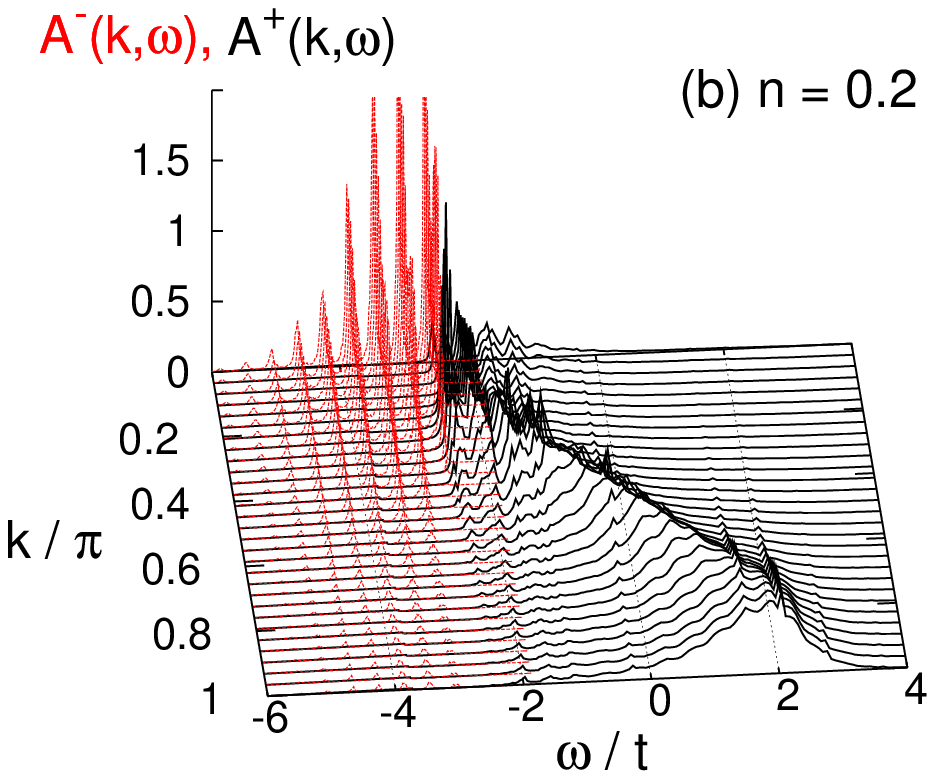}\\
    \includegraphics[width=0.47\textwidth]{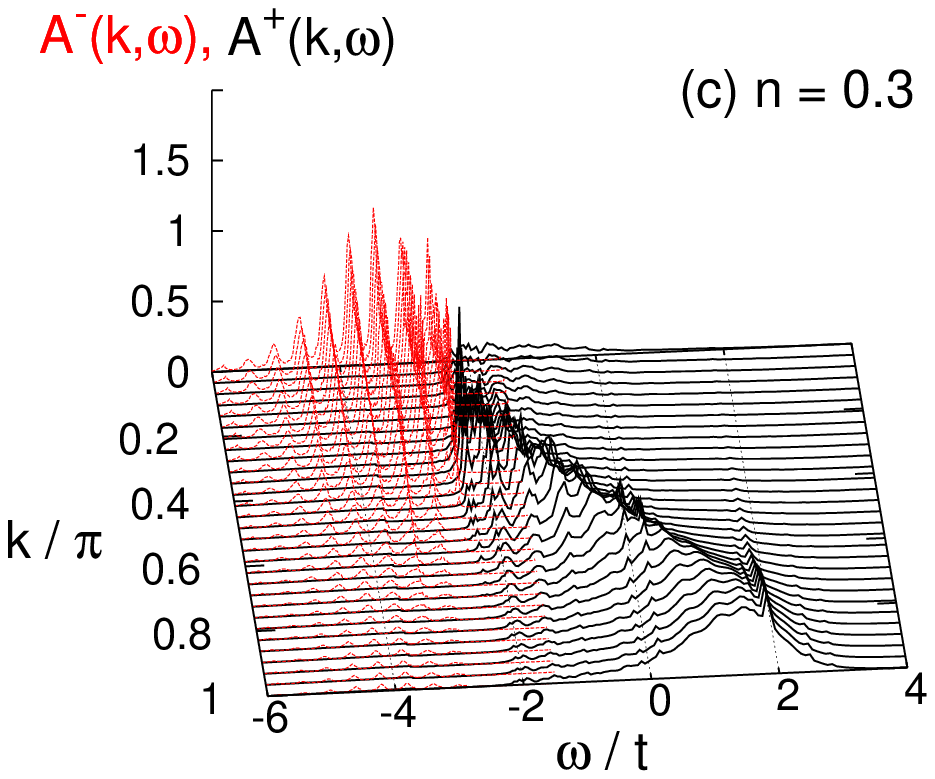}
    \includegraphics[width=0.47\textwidth]{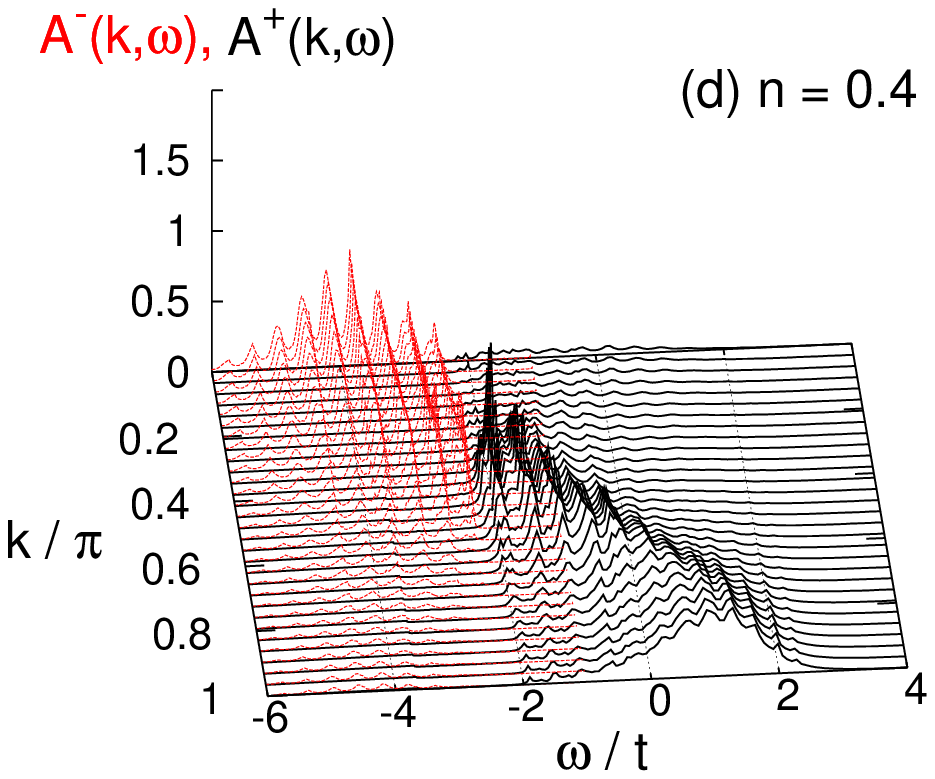}
    \caption{\label{fig:manypol_akw_density}
      (color online)
      One-electron spectral functions $A^-(k,\omega)$ (red dashed lines) and
      $A^+(k,\omega)$ (black solid lines) for different band fillings
      $n$. Here $\omega_0/t=0.4$, $N=10$ and $\lambda=1$. (a) $n=0.1$; (b)
      $n=0.2$; (c) $n=0.3$; (d) $n=0.4$.}    
\end{figure*}

A substantial dependence of the system properties on the band filling is
expected for intermediate electron-phonon interaction, for which a large
(extended) polaron state is known to exist in the single-electron case. By
contrast, for strong coupling, the polaron size collapses to a single site,
so that different carriers will not overlap (interact). Hence, we take
$\lambda=1$, the critical coupling for the large-to-small polaron cross-over
in the adiabatic regime $\omega_0/t<1$. Note that the abovementioned density
effects are absent for $\omega_0/t\gg1$ \cite{HoNevdLWeLoFe04}.

Figure~\ref{fig:manypol_akw_density} shows results for the one-electron
spectral functions $A^\pm(k,\omega)=-\mathrm{Im}\,G^\pm(k,\omega)/\pi$, where
$G^-$ ($G^+$) corresponds to the (inverse) photoemission part of $G$, for
fixed cluster size $N=10$.

In the low density case $n=0.1$ (Fig.~\ref{fig:manypol_akw_density}(a)), we
can easily identify a (coherent) polaron band crossing the Fermi level
$E_\mathrm{F}$, the latter being situated at the point where $A^-$ and $A^+$
intersect. The band flattens at large $k$, as known from single-polaron
studies. Below
this band, there exist equally spaced phonon excitations with small spectral
weight, which reflect the Poisson distribution of phonons in the one-electron
ground state.  Finally, above $E_\mathrm{F}$, there is a broad incoherent
band whose maximum follows closely the dispersion relation of free particles.

As the density $n$ increases
(Figs.~\ref{fig:manypol_akw_density}(a)\,--\,(c)) , the gap between the
polaron band and the incoherent excitations is reduced, and the low-energy
phonon peaks begin to broaden, until they have ultimately merged into a broad
band for $n=0.4$ (Fig.~\ref{fig:manypol_akw_density}(d)). More importantly,
for $n\geq0.3$, a polaron band can no longer be identified as
incoherent excitations lie very close to the Fermi level.  Finally, at
$n=0.4$, the spectrum consists of a broad main band of effective width
$\approx 6t$ crossing $E_\mathrm{F}$, as in a metallic system, and low-energy
excitations are available. This indicates that the polaronic quasiparticles,
existing at small $n$, have dissociated into dressed electrons due to mutual
interaction. Therefore, the system can no longer be described in terms of
small-polaron theory, as expected for intermediate electron-phonon coupling
at large $n$.

To summarize, the density-driven cross-over from a polaronic system to one
with moderately dressed electrons has been studied by means of CPT, whose
unlimited momentum resolution significantly simplifies the interpretation
compared to previous calculations \cite{HoNevdLWeLoFe04}. Finally, the
improvement of the cluster diagonalization represents an important step
toward future studies of even more complex problems.

This work was supported by the FWF project P15834, the DFG through SPP1073,
and KONWIHR. M.~H. is grateful to HPC-Europa.

\end{document}